\documentclass[a4paper]{jpconf}
\usepackage{graphicx}
\usepackage{amssymb}
\begin{document}
\title{Discovery potential for Higgs bosons beyond the SM}

\author{Gabriella P\'asztor \textit{for the CMS Collaboration}}

\address{University of California, Riverside, U.S.A. \\
KFKI Research Institute for Particle and Nuclear Physics, Budapest, Hungary}

\ead{Gabriella.Pasztor@cern.ch}

\begin{abstract}
The discovery potential of the CMS detector for the MSSM neutral and charged
Higgs bosons at the LHC is presented based on studies with full detector
simulation and event reconstruction of the principal discovery channels.
\end{abstract}

\section{Higgs bosons in the MSSM} 

In the Minimal Supersymmetric Standard Model (MSSM) the electroweak symmetry is
broken via the Higgs mechanism, leading to five physical scalars: three neutral
(h, H, A) and two charged (H$^+$, H$^-$) Higgs bosons. At tree level, two
parameters, usually chosen to be the mass of the pseudo scalar Higgs boson
($m_\mathrm{A}$) and the ratio of the vacuum expectation values of the two
Higgs fields ($\tan\beta$), determine the masses and couplings in the Higgs
sector. Radiative corrections, chiefly coming from the top-stop and, at large
$\tan\beta$, from the bottom - sbottom sector, introduce further model
parameters.

In the frequently studied decoupling limit ($m_\mathrm{A} \gg m_\mathrm{Z}$),
the lightest neutral Higgs boson is Standard Model-like, while H and A are
close in mass and both couple with $\tan\beta$ to down-type fermions and with
$\cot\beta$ to up-type fermions.

The dominant production mechanism for heavy neutral MSSM Higgs bosons at large
$\tan\beta$ is gg $\rightarrow$ b\=bH/A with negligible contributions from
other processes. As the main decay mode H/A $\rightarrow$ b\=b suffers from
high QCD background, the decay H/A $\rightarrow$ $\tau\tau$, reaching about
10\% rate for $\tan\beta>$10, promises the most sensitivity.

Charged Higgs bosons couple to fermions as H$^+$ud $\sim$ 
$m_\mathrm{d} \tan\beta + m_\mathrm{u} \cot\beta$. 
They are produced in top quark decay, if $m_\mathrm{H^\pm} <
m_\mathrm{t}$, or in association with a top quark,
if  $m_\mathrm{H^\pm} > m_\mathrm{t}$. Light charged
Higgs bosons decay almost exclusively to a tau lepton and a neutrino for
$\tan\beta\gtrsim 3$. For 
$m_\mathrm{H^\pm} > m_\mathrm{t}+m_\mathrm{b}$, the decay H$^\pm$ $\rightarrow$ tb dominates
with an important contribution of H$^\pm$ $\rightarrow$ $\tau \nu_\tau$ at
large $\tan\beta$.

The following processes are considered here:
\begin{itemize}
\item associated b\=bH/A production followed by H/A $\rightarrow$ $\tau\tau$
with all possible $\tau\tau$ final states (jet jet, e jet, $\mu$ jet and e
$\mu$)~\cite{bbH};
\item production of a light charged Higgs boson in t\=t
$\rightarrow$ H$^\pm$bWb with subsequent H$^\pm$ $\rightarrow$ $\tau
\nu_\tau$, $\tau$ $\rightarrow$
jet and W $\rightarrow$ $\ell\nu_\ell$ decays~\cite{lightHp};
\item associated H$^\pm$t(b) production of heavy charged Higgs boson with 
H$^\pm$ $\rightarrow$ $\tau\nu_\tau$, $\tau$ $\rightarrow$ jet and hadronic top quark
decays~\cite{heavyHp};
\item associated H$^\pm$t(b) production of heavy charged Higgs boson with 
H$^\pm$ $\rightarrow$ tb and one of the top quarks decaying
leptonically~\cite{hadrHp}.
\end{itemize}

\section{Experimental tools}

The final states in the search for heavy MSSM Higgs
bosons are complex: they typically contain
leptons, b- and light-flavoured hadronic jets and missing transverse energy. 

The selections rely on $\tau$-jet or lepton (e, $\mu$) triggers.
$\tau$-identification based on vertex reconstruction and impact parameter
measurements is powerful against hadronic jets. A single b-tag is sufficient to
suppress events from Drell-Yan, QCD multi-jet and W+jets processes. Against the
difficult t\=t background a central jet veto is applied. Missing transverse
energy ($E_\mathrm{T}^\mathrm{miss}$) reconstruction is important to account
for the neutrinos from the Higgs or the $\tau$ decays. Top quark and W boson
mass reconstruction provides a further handle on the background. For example,
in the bbH/A, H/A $\rightarrow$  $\tau\tau$ $\rightarrow$ $\ell$ + jet
searches, it is instrumental in vetoing leptons from W decay by reconstructing
the transverse mass of the [lepton, $E_\mathrm{T}^\mathrm{miss}$] system. 

The main background comes from t\=t events, with significant contribution from
W+jets events in the H$^\pm$ $\rightarrow$ $\tau\nu_\tau$ searches and Drell-Yan
processes in the H/A $\rightarrow$ $\tau\tau$ $\rightarrow$ $\ell$ + jet
searches. A major exception is the  H/A $\rightarrow$ $\tau\tau$ $\rightarrow$
jet jet analysis, where QCD events dominate the background.
The typical selection efficiency is below 1\%.

\medskip 

\noindent \textit{Higgs boson mass reconstruction} does not only increases the
sensitivity of the searches, but it will also play a crucial role in
constraining the MSSM model parameters. 

In the H/A $\rightarrow$ $\tau\tau$ selection the Higgs mass can be measured
by the di-tau mass assuming that the $\tau$ decay products are
collinearly emitted. While the procedure has large inefficiency, it achieves a
mass resolution of about 20\%.

In the heavy charged Higgs search: H$^\pm$t $\rightarrow$  $\tau\nu_\tau$ bqq,
the reconstruction of the top quark mass is instrumental in fully
reconstructing the signal. The charged Higgs mass is then estimated by the
transverse mass of the [$\tau$, $E_\mathrm{T}^\mathrm{miss}$] system
($M_\mathrm{T}$). By requiring  $M_\mathrm{T}>$100~GeV, an almost
background-free selection is achieved.

\section{Results}

The results presented here are based on the full simulation of the CMS detector
at low luminosity (${\cal L}=2\cdot 10^{33}$~cm$^{-2}$s$^{-1}$), assuming a
total integrated luminosity of 30$-$60 fb$^{-1}$. Pythia is used as the main
Monte Carlo generator with notable exceptions (e.g. matrix element generators
for multi-parton or Toprex for t\=t final states in some cases). Background
cross-sections are normalized to NLO calculations where available. Tau decays
are modeled by Tauola. 

For the numerical results, the minimal supergravity inspired $m_\mathrm{h}$-max
benchmark scenario is used, which maximises the theoretical upper bound on
$m_\mathrm{h}$ for a given $\tan\beta$, $m_\mathrm{t}$ and
$M_\mathrm{SUSY}$, the SUSY mass scale. Higgs boson masses, cross-sections and
branching ratios are calculated by FeynHiggs 2.3.2. 

\begin{figure}[ht!]
\includegraphics[width=12.5pc,height=13.5pc]{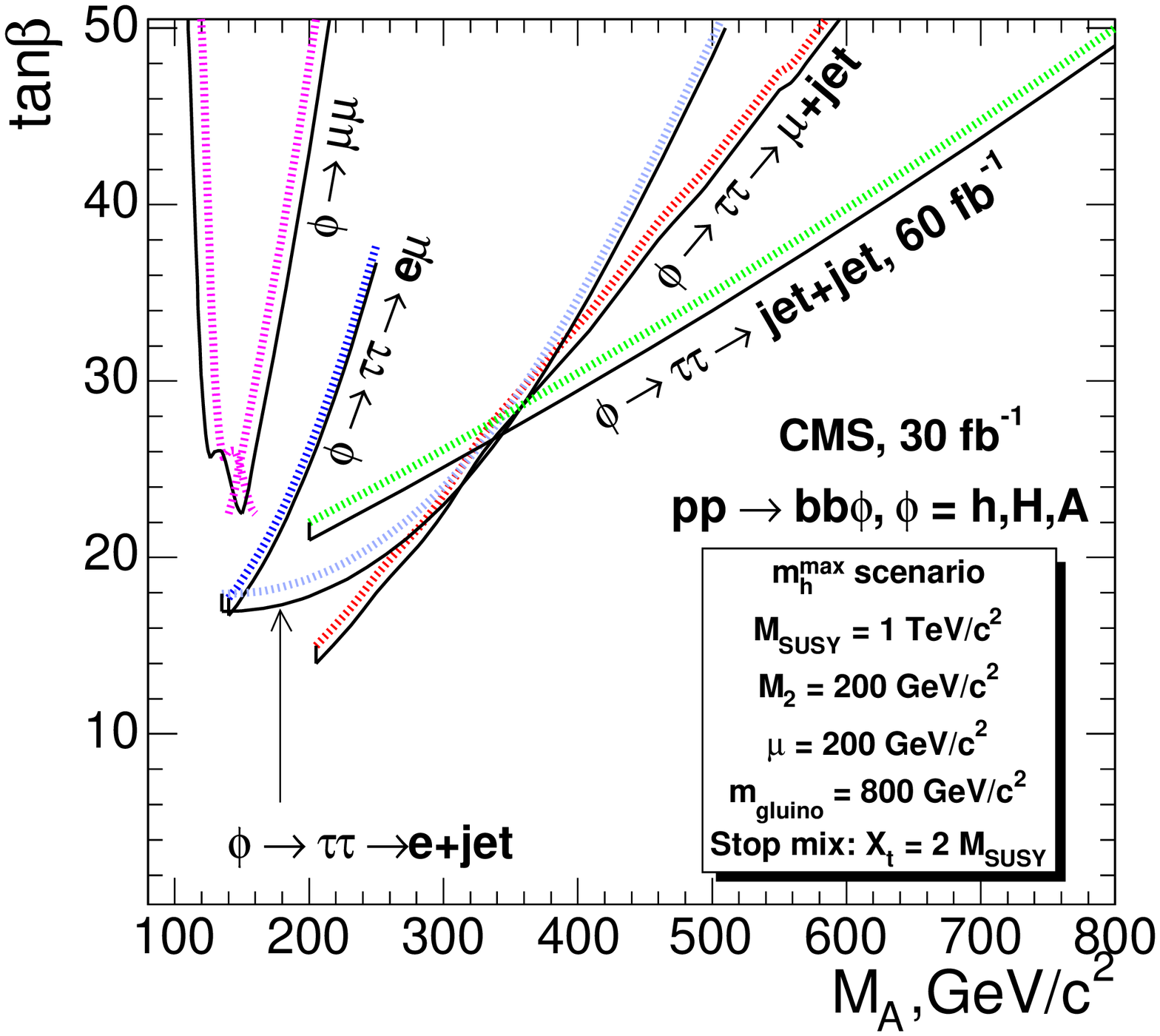}
\includegraphics[width=12.5pc,height=13.5pc]{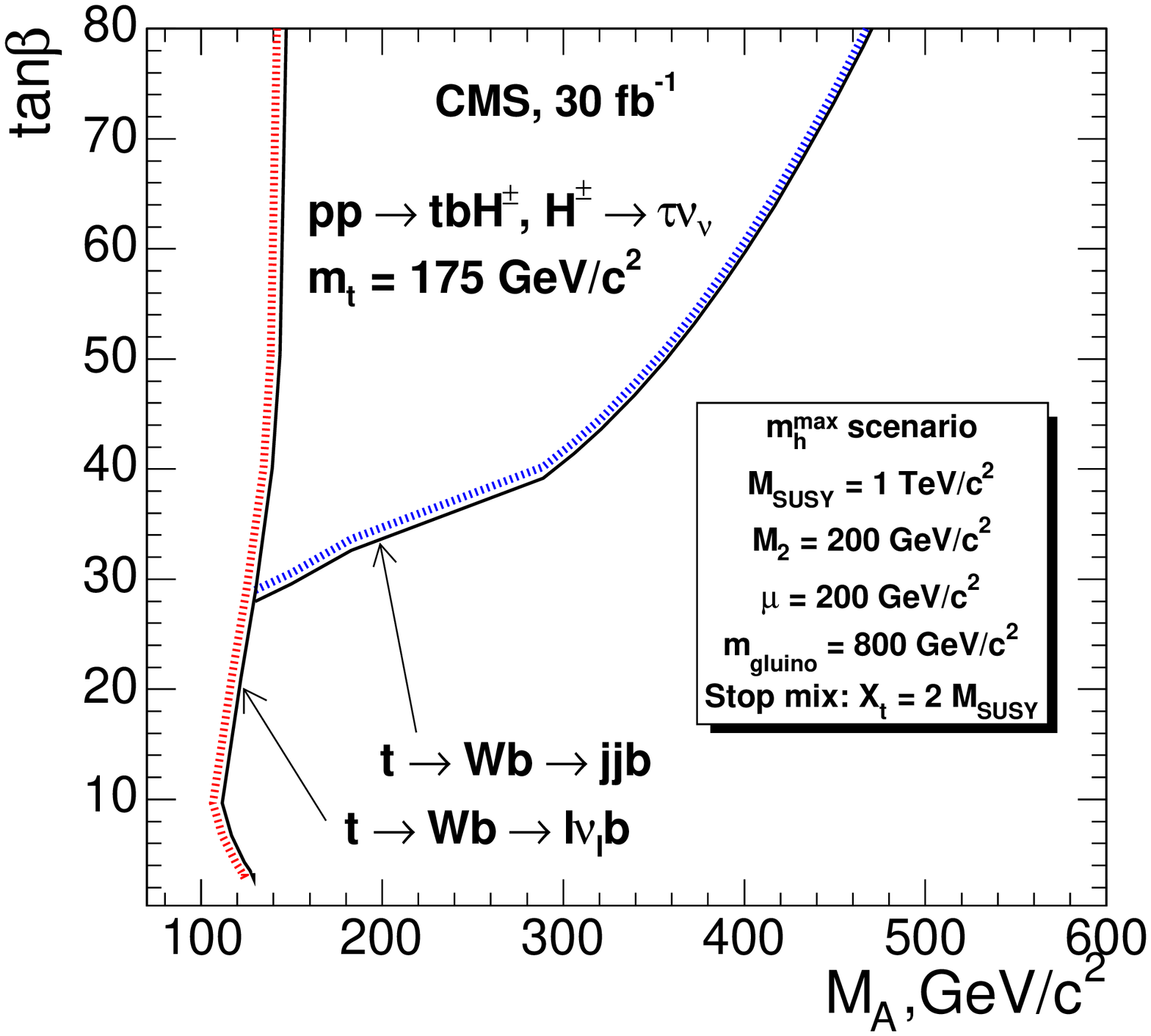}
\includegraphics[width=12.5pc,height=13.5pc]{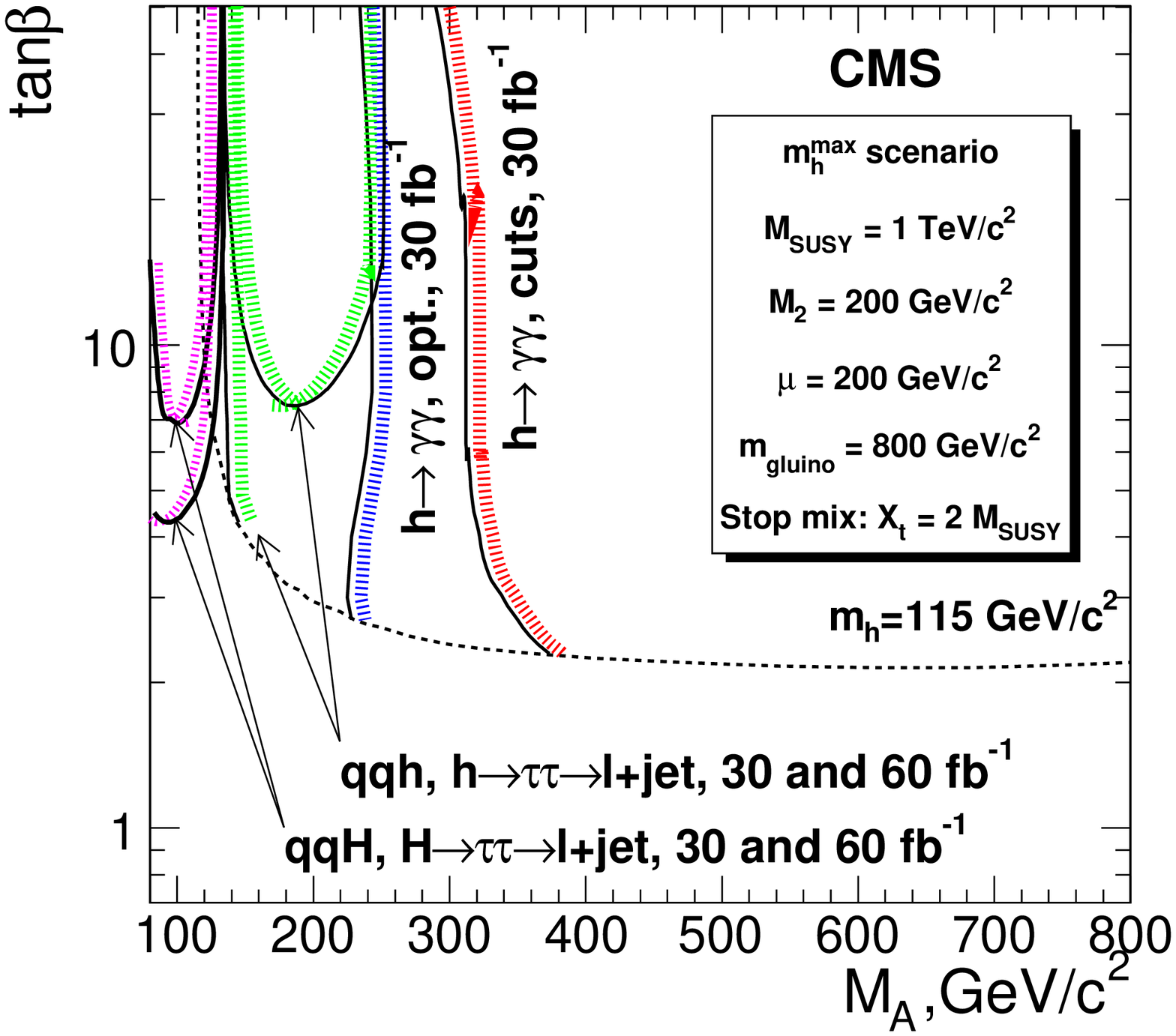}

\vspace*{-5.43cm}

\hspace*{4.3cm} (a) \hspace*{10.9pc} (b) \hspace*{11pc} (c)

\vspace*{4.9cm}

\caption{\label{fig:h} 5-sigma discovery reach of searches for 
(a) bbH/A production with H/A $\rightarrow$ $\tau\tau$, 
(b) H$^\pm$ $\rightarrow$ $\tau\nu_\tau$ and 
(c) inclusive h $\rightarrow$ $\gamma\gamma$ and 
vector boson fusion qqh/H with h/H $\rightarrow$ $\tau\tau$.
}
\end{figure}

The estimated 5-sigma discovery reach is shown on Figure~\ref{fig:h}(a) for the
pp $\rightarrow$ bbH/A searches and on Figure~\ref{fig:h}(b) for the H$^\pm$
$\rightarrow$ $\tau\nu_\tau$ searches~\cite{PTDR}. In all cases the systematic
uncertainties are included in the calculation of signal significance. 

We also considered the extended $m_\mathrm{h}$-max scenario, where the
supersymmetric Higgs mass parameter can take different values: $\mu = \pm
200, \pm 500, \pm 1000$. The cross-section is enhanced for large negative
and reduced for large positive values of $\mu$, but the change in the Higgs
branching ratios due to decay modes to supersymmetric particles partially
compensates this effect. In general,
the default value of $\mu=200$~GeV gives the most conservative discovery reach.
For a detailed discussion see~\cite{mu-dep}. 

\medskip

\noindent \textit{Systematic uncertainties} dominate these searches. Therefore,
several methods are explored to measure them from collision
data~\cite{syst-estim}.  

The main sources of uncertainty come from b-tagging, $\tau$- and lepton
identification, missing energy measurement and energy scale calibration.
Together with the theoretical uncertainties on the background production
cross-sections, we estimate typically 12\% error on the main t\=t background,
9\% on Z/$\gamma$, 16\% on bbZ/$\gamma$, 10$-$14\% on W+jets and 15\% on tW
processes. The QCD background will be measured from data with a 5$-$20\%
statistical error. 

The inclusion of systematic uncertainties has a significant impact on the
expected sensitivities. For example, in the charged Higgs boson H$^\pm$
$\rightarrow$ $\tau \nu_\tau$ search, the 5-sigma reach is decreased from
about 125 to 110~GeV at the most difficult value of $\tan\beta\approx 10$. At
values of $\tan\beta>30$, the downward shift is even larger ranging from
about 30 to 80~GeV.

In the search for gg $\rightarrow$ H$^\pm$tb with H$^\pm$ $\rightarrow$ tb and
tt $\rightarrow$ $\ell\nu_\ell$b qqb requiring four b-tags, the main background
comes from ttbb and mistagged tt + jets processes. With the simulation of
the background processes by CompHEP (with the cross-section calculated by
ALPGEN) and the inclusion of realistic experimental (about 5$-$5\% on b-tagging
efficiency and the mistag rate) and theory uncertainties, no discovery
potential remains for this channel.

\section{Light neutral Higgs search in MSSM}

The searches presented above for heavy Higgs bosons loose their sensitivity for
small and intermediate $\tan\beta$ and leave open the so-called LHC wedge
region, where only a light neutral Higgs boson can be discovered. This is
illustrated on figure~\ref{fig:h}(c), where the SM Higgs boson
searches~\cite{PTDR,sm} (for inclusive pp $\rightarrow$ h+X production with h
$\rightarrow$ $\gamma\gamma$ and for the vector boson fusion process qq
$\rightarrow$ qqh/H with h/H $\rightarrow$ $\tau\tau$ $\rightarrow$ $\ell$ +
jet) are reinterpreted in MSSM~\cite{PTDR}. 

\ack

The author is partially supported by OTKA Grant NK67974.

\end{document}